\documentclass[10pt]{article}
\usepackage{graphicx}
\usepackage{amsmath}
\usepackage{amsthm}
\usepackage{amssymb}
\usepackage{amsfonts}
\usepackage{bbm}
\usepackage{url}
\usepackage{authblk}
\usepackage{enumitem}
\usepackage[a4paper, margin=1in]{geometry}

\newcommand{\ket}[1]{| #1\rangle}
\newcommand{\bra}[1]{\langle #1 |}
\newcommand{\ketbra}[1]{\ket{#1}\bra{#1}}
\newcommand{\tr}{{\rm tr}}

\newcommand{\G}{\mathcal{G}}

\newcommand{\E}{\mathbb{E}}

\newcommand{\cU}{{\mathcal U}}
\newcommand{\cZ}{{\mathcal Z}}
\newcommand{\cH}{{\mathcal H}}
\newcommand{\cW}{{\mathcal W}}
\newcommand{\cT}{{\mathcal T}}
\newcommand{\cO}{{\mathcal O}}

\renewcommand{\epsilon}{\varepsilon}

\newcommand{\qa}{\alpha}

\newcommand{\qe}{\varepsilon}
\newcommand{\qs}{\sigma}
\newcommand{\qg}{\gamma}

\newcommand{\qf}{\varphi}
\newcommand{\ql}{\lambda}
\newcommand{\qL}{\Lambda}
\newcommand{\qt}{\tau}

\newcommand{\qr}{\rho}
\newcommand{\qJ}{\Psi}
\newcommand{\qF}{\Phi}

\newcommand{\vece}{{\bf e}}

\newcommand{\fr}[2]{{\textstyle \frac{#1}{#2}}}
\newcommand{\isdef}{\stackrel{\rm def}{=}}
\newcommand{\bits}{ \{ 0,1\}  }

\newcommand{\be}{\begin{equation}}
\newcommand{\ee}{\end{equation}}
\newcommand{\bea}{\begin{eqnarray}}
\newcommand{\eea}{\end{eqnarray}}

\renewcommand{\S}{\mathcal{S}}

\newtheorem{theorem}{Theorem}[section]

\newtheorem{lemma}[theorem]{Lemma}

\title{A simpler security proof for 6-state quantum key distribution}
\author[1]{Kaan Akyuz}
\author[2]{Boris \v{S}kori\'{c}}
\affil[1]{\small Middle East Technical University, Turkey}
\affil[2]{\small TU Eindhoven, The Netherlands}
\date{}

\begin{document}

\setlength{\parindent}{0mm}

\maketitle

\begin{abstract}
\noindent
Six-state Quantum Key Distribution (QKD) achieves the highest key rate in the class of qubit-based QKD schemes.
The standard security proof, which has been developed since 2005, invokes complicated theorems involving smooth R\'{e}nyi entropies.

\noindent
In this paper we present a simpler security proof for 6-state QKD that entirely avoids R\'{e}nyi entropies.
This is achieved by applying state smoothing directly in the Bell basis.
We furthermore show that the same proof technique can be used for 6-state quantum key recycling.
\end{abstract}

%==========================================================================
\section{Introduction}
\label{sec:intro}

Early security proofs  for quantum key distribution \cite{Mayers1996,BBBMR2000,Lo2001,GottesmanLo2003,Inamori2000,SP2000}
were not formulated in the universal composability framework.
The universal composability approach 
has been followed for QKD since 2005 
\cite{renner2008security,BHLMO2005,KGR2005,RGK2005,RK2005}.
This has led to security proofs in which the Leftover Hash Lemma (LHL) against quantum adversaries
plays a central role. The LHL provides an upper bound on the distinguishability between the generated QKD key
and a completely random string, given all classical and quantum information held by the adversary.
All the various versions of the LHL work with smooth R\'{e}nyi entropies \cite{KRS2009,MDSFT2013}
and invoke theorems about their properties.
Hence, reading a QKD security proof requires an understanding of rather advanced concepts
and a heavy theoretical toolbox.

In this paper we provide a more `schoolbook' security proof for 6-state QKD that entirely avoids R\'{e}nyi entropies.
We rely on postselection \cite{CKR2009} to lift security against collective attacks to security against general attacks.
We follow a number of steps familiar from the LHL, but at the point where one would usually
rewrite expressions in terms of R\'{e}nyi entropies we work with expressions that are diagonalized
in the Bell basis, so that square roots of operators can be explicitly computed.
We apply smoothing (cutting off probability tails) in the Bell basis, in a way that resembles
smoothing of classical probability distributions.
This yields a finite-size result for the key rate, with $\cO(1/\sqrt n)$ finite-size contributions,
which is the same order that the standard security proof gives.

We focus on 6-state QKD for several reasons:
(i)
Among qubit-based QKD schemes
it stands out as the one with the highest key rate as a function of the quantum bit error rate (QBER).
(ii)
For BB84 very powerful proofs exist that immediately yield general security without needing to 
go via collective attacks and postselection. 
These do not work for the high rate of 6-state QKD.
(iii)
The level of simplification that our proof provides is more compelling for 6-state than for BB84. 

The outline of the paper is as follows.
In the preliminaries (Section~\ref{sec:prelim}) we briefly review the standard security proof for
6-state QKD, and we list a number of lemmas that we will use.
We present our simplified proof in Section~\ref{sec:proofQKD}, 
and we plot key rates as a function of QBER for various finite sizes, 
showing convergence to the asymptotic rate.
In Section~\ref{sec:discussion} we discuss possible improvements.
In the Appendix we show that the proof technique can also be applied to
6-state quantum key recycling.

%---------------------------------------------------------------------
\section{Preliminaries}
\label{sec:prelim}

\subsection{Notation}
Classical Random Variables are denoted with capital letters, and their realisations with lowercase letters. 
Sets are denoted in calligraphic font. 
The probability that $X$ takes value $x$ is written as $\rm Pr[X=x]$. 
The expectation with respect to $X$ is denoted as $\mathbb{E}_x f(x)=\sum_{x \in \mathcal{X}} {\rm Pr}[X=x] f(x)$. 
The notation 'log' stands for the logarithm with base 2. 
%The Kronecker delta is denoted as $\delta_{a b}$. 
We write the binary entropy function as $h(p)=p\log\fr1p+(1-p)\log\fr1{1-p}$, and more generally
$h(p_1,\ldots,p_N)=\sum_{i=1}^N p_i \log\fr1{p_i}$.
The inverse of a bit $b \in\{0,1\}$ is $\bar{b}=1-b$. 
The Hamming weight of a string $x$ is written as $w(x)=| \{ i: w_i\neq 0 \} |$. 
%We will speak about the bit error rate $\gamma$ of a quantum channel. 
%This is defined as the probability that a classical bit $g$, sent by Alice embedded in a qubit, arrives at Bob's side as $\bar{g}$. 
We write $\mathbb{I}$ for the identity matrix. 
The notation $\operatorname{tr}$ stands for trace. 
The Hermitian conjugate of an operator $A$ is written as $A^{\dagger}$. 
%The complex conjugate of a scalar $z$ is $z^*$. 
Let $A$ have eigenvalues $\lambda_i$. 
The 1-norm of $A$ is written as $\|A\|_1=\operatorname{tr} \sqrt{A^{\dagger} A}=\sum_i\left|\lambda_i\right|$. 
$\mathcal{S}(\mathcal{H})$ denotes the space of positive semidefinite
operators on the Hilbert space $\mathcal{H}$. 
The trace distance between operators $\rho,\sigma$
is $\|\rho-\sigma\|_{\rm tr}  =\frac{1}{2}\|\rho-\sigma\|_1$.

%---------------------------------------------------------------------
\subsection{`Standard' security proof for 6-state QKD}
\label{sec:prelim6state}

We briefly review the security analysis for 6-state QKD 
with a single-photon source,
with one-way classical postprocessing and without artificial preprocessing noise.
We focus on the proof technique developed by Renner et al. \cite{renner2008security,KGR2005,RGK2005,RK2005,CKR2009}, 
which yields the highest key rate
while satisfying universal composability \cite{BHLMO2005,RK2005}. 
An important ingredient is the use of {\em post-selection}  \cite{CKR2009},
which makes it possible to upgrade a security proof in case of collective attacks
to a security proof in case of general attacks.\footnote{
More recent techniques based on entropic uncertainty relations \cite{TLGR2012,TL2017} do not need such a step and immediately
yield finite-size results for any attack. 
However, they do not work for the high rates of 6-state QKD.
}
The cost of this upgrade is a modest reduction of the key length, by $30\log(n+1)$ bits.
A second main ingredient is {\em symmetrisation} \cite{renner2008security,RGK2005}. 
Alice and Bob share $n$ noisy EPR pairs.  
The security of the protocol does not change if they both apply the same
Pauli operations on their own qubits, chosen at random independently for each EPR pair.
The joint effect of postselection and symmetrisation is that 
it suffices to consider states of the form $(\qs^{AB})^{\otimes n}$,
where $\qs^{AB}$ is a two-qubit density matrix that is diagonal in the Bell basis
and depends only on the QBER.
For states $(\qs^{AB})^{\otimes n}$ that successfully pass the parameter estimation step
of the QKD protocol, we may write
\be
    \qs^{AB}=(1-\fr32\qg)\ketbra{\qJ^-} + \fr\qg2 \ketbra{\qF^-}+ \fr\qg2 \ketbra{\qJ^+}  
    + \fr\qg2 \ketbra{\qF^+}
\label{symmetrisedAB}
\ee
where $\qg$ is the maximum allowed QBER.
(Here we have taken the EPR pairs to be singlet states.)
As a worst-case assumption
it is considered that Eve holds the purification of the AB system.
Using notation similar to \cite{LS2018} one can write the purification as
\be
    \ket{\qJ^{ABE}} = \sqrt{1-\fr32\qg}\ket{\qJ^-}\ket{0}
    +\sqrt{\fr\qg2}\Big( -\ket{\qF^-}\ket{1} +i\ket{\qJ^+}\ket{2} +\ket{\qF^+}\ket{3} \Big)
\ee
which leads to a simple form for Eve's post-measurement state.
Alice and Bob do a measurement in a basis that is characterised by spin direction 
%$\vecv=(v_1,v_2,v_3)$ 
$\vece_j$
on the Bloch sphere, where $j\in\{1,2,3\}$ stands for the $x,y,z$-axis respectively.
Alice's outcome is $x\in\bits$ and Bob's outcome is $y\in\bits$.
Eve's post-measurement state, conditioned on outcomes $x,y$, is  
$\qs^j_{xy} = \ketbra{E^j_{xy}} $
with
\bea
    \ket{E^j_{x\bar x}} &=& \frac1{\sqrt{1-\qg}} \Big[ \sqrt{1-\fr32\qg}\ket{0}+(-1)^x\sqrt{\fr\qg2} \ket{j} \Big]
\label{defE01}
%    \\ 
%    \ket{E^j_{10}} &=& \frac1{\sqrt{1-\qg}} \Big[ \sqrt{1-\fr32\qg}\ket{0}-\sqrt{\fr\qg2} \ket{j} \Big]
    \\
    \ket{E^j_{xx}} &\propto&  \frac1{\sqrt2} \Big[\ket{j+1}+i(-1)^{x+1} \ket{j+2} \Big]
%    \\
%    \ket{E^j_{11}} &\propto&  \frac1{\sqrt2} \Big[\ket{j+1}+i \ket{j+2} \Big]
\label{defE11}
\eea
where the indices $j+1$, $j+2$ are understood to cycle back into $\{1,2,3\}$.
(This state of Eve is also obtained from optimal attacks analysis \cite{SKMB2008}.)
The full post-measurement state is
\bea
    \qr^{JXYE} &=& \sum_{j\in\{1,2,3\}^n}{\rm Pr}[J=j]\sum_{x,y\in\bits^n} p_{xy}\ketbra{j,x,y}\otimes
    \qr^E_{jxy}
    \\
    p_{xy} &=& 2^{-n}\qg^{w(\bar x\oplus y)}(1-\qg)^{n-w(\bar x\oplus y)}
    \\
    \qr^E_{jxy} &=&
    \bigotimes_{i=1}^n \qs^{j_i}_{x_i y_i}.
\eea
Alice sends the syndrome of $x$ to Bob, one-time-pad encrypted.
This allows Bob to reconstruct $x$ from $y$ and the syndrome.
(If the reconstruction fails then Alice and Bob abort.)
The QKD key $z\in\cZ$ is derived from $x$ as $z=\qF(u,x)$, where $\qF$ is a universal hash function and
$u\in\cU$ is a public seed.
The security proof amounts to upper bounding the statistical distance (trace distance) between
on the one hand
$Z$ given all of Eve's information and on the other hand a uniform variable on $\cZ$.
The encrypted syndrome does not enter into this analysis since the one-time pad key is
entirely independent; the sending of this ciphertext ends up only as a penalty term
in the QKD key rate due to the expenditure of key material.
The quantity to be upper bounded is
\be
    D = \|  \qr^{Z UJ E}-\mu^Z\otimes \qr^{UJE} \|_{\rm tr}.
\label{defDtracedist}
\ee
It can be written as
$D=\fr12 \tr \sum_j {\rm Pr}[J=j]\sum_{z,u}\frac1{|\cU|}\sqrt{(p_{z|u} \qr^E_{jzu} - \frac1{|\cZ|} \qr^E_{ju})^2} $.
The first step is to pull the sums $\sum_{zu}$ into the square root with a Jensen inequality,
and make use of the universal hash properties to evaluate these sums.
The result is
$D\leq\fr12 \sum_j {\rm Pr}[J=j] \tr \sqrt{|\cZ|  \sum_x p_x^2 (\qr^E_{jx})^2 }$.
However, Jensen's inequality is so un-tight that it pays off to take a different starting point
before applying the inequality.
A {\em smoothed} state $\bar\rho$ is considered, which lies close to $\rho$.
It holds that
\bea
    D &\leq& 2 \| \qr^{Z UJ E} - \bar\qr^{Z UJ E}  \|_{\rm tr} + \bar D
\label{Dtriangle}
    \\
    \bar D &\isdef& \|  \bar\qr^{Z UJ E}-\mu^Z\otimes \bar\qr^{UJE} \|_{\rm tr}
   \leq  \fr12 \sum_j {\rm Pr}[J=j] \tr \sqrt{|\cZ|  \sum_x p_x^2 (\bar\qr^E_{jx})^2 }.
\label{defDbar}
\eea
Next the trace too is pulled into the square root with Jensen, which yields an extra factor
support$(\qr^E_j)$ inside the square root.
Then it is noted that the expression $\log\sum_x p_x^2 (\bar\qr^E_{jx})^2$
is a R\'{e}nyi 2-entropy, whereas log(support$(\bar\qr^E_j))$ is a R\'{e}nyi 0-entropy.
Finally a number of `sledgehammer' theorems are invoked
to bound entropies of $\bar\qr$ by smooth entropies of $\qr$ \cite{RK2005,renner2008security} and finally
to bound the smooth R\'{e}nyi entropies by von Neumann entropies \cite{renner2008security},
in particular the von Neumann entropy of the averaged state $\qr^E=\sum_{jxy}p_j p_{xy}\qr^E_{jxy}$
which is identical in form to $\qs^{AB}$ (\ref{symmetrisedAB}).
The end result is that asymptotically
$\bar D\leq \fr12\E_j \sqrt{|\cZ| 2^{-n} 2^{S(E|j)-S(E|Xj) } }$
$=\fr12\sqrt{2^{\ell-n} 2^{n h(1-\fr32\qg,\fr\qg2,\fr\qg2,\fr\qg2) -n h(\qg) } }$.
(Here $S$ stands for von Neumann entropy, and we have written $|\cZ|=2^\ell$.)
Hence the QKD key length $\ell$ can be set to slightly below
$n-n h(1-\fr32\qg,\fr\qg2,\fr\qg2,\fr\qg2) +n h(\qg)$.
Taking into account the key material spent on sending the syndrome, which asymptotically has size $nh(\qg)$,
the asymptotic key rate is given by $\frac1n[\ell -nh(\qg)]$,
\be
    \mbox{6-state QKD asymptotic key rate = }1-h(1-\fr32\qg,\fr\qg2,\fr\qg2,\fr\qg2).
\label{asymprate}
\ee

%---------------------------------------------------------------------
\subsection{Useful Lemmas}

\begin{lemma}[Lemma A.2.8 in \cite{renner2008security}]
\label{lemma:distbound}
Let $\rho,\bar\rho\in{\cal P}(\cal H)$ with $\bar\rho=P\rho P$ for some projector $P$ on~$\cal H$. 
Then 
\begin{align}
    \|\rho-\bar{\rho}\|_{1} \leq 2 \sqrt{\operatorname{tr}\rho\, \operatorname{tr}(\rho-\bar{\rho})}. 
\end{align}
\end{lemma}

\begin{lemma}
\label{lemma:Bretagnolle}
(Bretagnolle–Huber–Carol inequality. Proposition~2 in \cite{BH1978}.)
Let $(Z_1,\ldots,Z_t)$ be a multinomial-distributed vector with parameters 
$(\pi_1,\ldots,\pi_t)$, satisfying $\sum_{s=1}^t Z_s =n$.
Then
\be
    {\rm Pr}\left[ \sum_{s=1}^t |Z_s-n\pi_s| \geq \qa\sqrt n  \right] \leq 
    2^t e^{-\frac12 \qa^2}.
\ee
\end{lemma}

%============================================================================
\section{Simplified security proof for 6-state QKD}
\label{sec:proofQKD}

%-------------------------------------------------
\subsection{Diagonal form in the Bell basis}

We present a relatively simple security proof for 6-state QKD that uses smoothing but
avoids R\'{e}nyi entropies altogether.
We take advantage of postselection and symmetrisation just like the proof discussed in
Section~\ref{sec:prelim6state}.
The point where we start to depart from the standard approach
is (\ref{Dtriangle},\ref{defDbar}).
We note that the expression $A_j\isdef \sum_x p_x^2 (\qr^E_{jx})^2$
is diagonal in the Bell basis.
We apply a smoothing procedure that acts as a projection $P_\S$ onto a subspace of
Eve's Hilbert space $\cH_E^{\otimes n}$.
We choose this subspace such that $P_\S A_j P_\S $ is still diagonal.
We define the set $\G=\{0,1,2,3\}^n$.
For $g\in\G$ we define the state $\ket g\in\cH_E^{\otimes n}$ as
\be
    \ket g = \bigotimes_{i=1}^n \ket{g_i}.
\label{defgstate}
\ee

\begin{lemma}
\label{sumsigmaisdiag}
Let $j\in\{0,1,2,3\}$ and $r\in\bits$.
\be
    \frac12 \sum_{x\in\bits}  \qs^j_{x,\bar x\oplus r} = \left\{\begin{array}{ll}
    r=0: & \frac{1-\fr32\qg}{1-\qg}\ketbra{0}+\frac{\qg/2}{1-\qg}\ketbra{j}
    \\
    r=1: & \frac12\ketbra{j+1} +\frac12\ketbra{j+2}
    \end{array}\right.
\ee
\end{lemma}
\underline{\it Proof:}
Follows directly from $\qs^j_{xy}=\ketbra{E^j_{xy}}$ with $\ket{E^j_{xy}}$ as given in (\ref{defE01}),(\ref{defE11}).
\hfill$\square$

\begin{lemma}
\label{lemma:Adiag}
Let $j\in\{1,2,3\}^n$ and $g\in\G$.
Let $t_0(g)=| \{  i: g_i=0 \} | $ be the tally of zeroes in~$g$.
Let $t_{\rm eq}(g,j) = | \{ i: g_i=j_i \}  | $ be the tally of places where $t$ and $j$ coincide.
Similarly, 
let $t_{+1}(g,j)  = | \{ i: g_i=j_i+1 \}  | $  and
$t_{+2}(g,j)  = | \{ i: g_i=j_i+2 \}  | $ where it is understood that $j+1$ and $j+2$
cycle back into the set $\{ 1,2,3 \}$.
Then it holds that
\bea
    \sum_x p_x^2 (\qr^E_{jx})^2 &=& \sum_{g\in\G} \ql_g(j) \ketbra{g} 
\label{diaggg}
    \\
    \ql_g(j) & = & 2^{-n}\Big[(1-\qg)(1-\fr32\qg)\Big]^{t_0(g)}  \Big[\fr\qg2(1-\qg)\Big]^{t_{\rm eq}(g,j)}
    \Big[ \fr12\qg^2 \Big]^{t_{+1}(g,j) + t_{+2}(g,j) }.
\label{eigenvalueA}
\eea
\end{lemma}
\underline{\it Proof:}
We have $\qr^E_{jx} = \bigotimes_{i=1}^n[ (1-\qg)\qs^{j_i}_{x_i\overline{x_i}} + \qg \qs^{j_i}_{x_i x_i} ]$.
Using the fact that the sigma matrices with $x=y$ are orthogonal to those with $x\neq y$ we get
$(\qr^E_{jx})^2 = \bigotimes_{i=1}^n[ (1-\qg)^2\qs^{j_i}_{x_i\overline{x_i}} + \qg^2 \qs^{j_i}_{x_i x_i} ]$.
Next we use $p_x=2^{-n}$ to obtain
$A_j=\sum_x p_x^2 (\qr^E_{jx})^2$ 
$=2^{-n} \bigotimes_{i=1}^n [(1-\qg)^2 \frac{\qs^{j_i}_{01}+\qs^{j_i}_{10}}2+\qg^2\frac{\qs^{j_i}_{00}+\qs^{j_i}_{11}}2]$.
Lemma~\ref{sumsigmaisdiag} tells us that this expression is diagonal in the Bell basis.
The eigenvectors are of the form (\ref{defgstate}).
We find the eigenvalues by computing $A\ket g$.
We see that every occurrence of $g_i=0$ generates a factor
$(1-\qg)^2 \frac{1-\fr32\qg}{1-\qg}=(1-\qg)(1-\fr32\qg)$.
Similarly, each occurrence $g_i=j_i$ yields a factor $(1-\qg)^2\frac{\qg/2}{1-\qg}=\fr\qg2(1-\qg)$.
Finally, $g_i\notin\{0,j_i\}$ leads to a factor $\qg^2\cdot\fr12$.
Counting how often each factor occurs yields (\ref{eigenvalueA}).
\hfill$\square$

If no smoothing is applied at all, Lemma~\ref{lemma:Adiag} directly yields a bound on 
the trace distance $D$~(\ref{defDtracedist}).

\begin{lemma}[Without smooting]
\label{lemma:withoutsmooth}
The distance 
$D = \|  \qr^{Z UJ E}-\mu^Z\otimes \qr^{UJE} \|_{\rm tr}$
for the state $\qr^E_{jx} = \bigotimes_{i=1}^n[ (1-\qg)\qs^{j_i}_{x_i\overline{x_i}} + \qg \qs^{j_i}_{x_i x_i} ]$ can be bounded as
\be
    D \leq \frac12\sqrt{2^{\ell-n}}
    \Big[ \sqrt{ (1-\qg)(1-\fr32\qg) } + \sqrt{\fr\qg2(1-\qg)} + 2\sqrt{\qg^2/2 } \Big]^n.
\label{Dnosmooth}
\ee
\end{lemma}
\underline{\it Proof:}
We substitute (\ref{diaggg}) into (\ref{defDbar}) without smoothing.
The resulting expression contains
$\tr\sqrt{\sum_x p_x^2 (\qr^E_{jx})^2}$
$=\sum_{g\in\G} \sqrt{\ql_g(j)} $.
Substituting  (\ref{eigenvalueA})
yields a summand that depends only on tallies.
The sum $\sum_{g\in\G}$ then simplifies to the form
$\sum_{\rm tallies}{n\choose{\rm tallies}}$ which is evaluated using the multinomial sum rule.
\hfill$\square$

Lemma~\ref{lemma:withoutsmooth} yields a rate that is decidedly worse than the standard result~(\ref{asymprate}).

%-----------------------------------------------------------------
\subsection{Explicit recipe for smoothing}

We pick a subset $\cT\subset\{ (a,b,c,d)\in {\mathbb N}^4 | a+b+c+d=n \}$.
This will represent the set of tallies that remain after smoothing.
We define sets
\be
    \S_j \isdef \{ g\in\G | \Big(t_0(g), t_{\rm eq}(g,j), t_{+1}(g,j), t_{+2}(g,j)  \Big)\in\cT  \}.
\ee
We introduce projection operators
\be
    P^j \isdef \sum_{g\in\S_j} \ketbra{g}.
\ee
For each combination of classical variables $(j,x,y)$ with $j\in\{1,2,3\}^n$
and $x,y\in\bits^n$ we apply smoothing as follows
\be
    \bar\qr^E_{jxy} = P^j \qr^E_{jxy} P^j.
\ee

\begin{lemma}
\label{lemma:epsilondist}
It holds that
\be
    \|  \qr^{ZUJE}-\bar\qr^{ZUJE} \|_{\rm tr} \leq  \sqrt{
    \sum_{(\qt_0,\qt_1,\qt_2,\qt_3)\notin\cT} {n\choose \qt_0,\qt_1,\qt_2,\qt_3} (1-\fr32\qg)^{\qt_0} (\fr\qg2)^{\qt_1+\qt_2+\qt_3}
    }.
\label{epsilondist}
\ee
\end{lemma}
\underline{\it Proof:}
The state $\qr^{ZUJE}$ (\ref{Dtriangle}) is given by
\be
    \qr^{ZUJE} = \sum_{zuj}p_j \frac1{|\cU|}\ketbra{z,u,j}\otimes \sum_{xy} p_{xy} p_{z|ux} \qr^E_{jxy}
\ee
and hence the smoothed version is
\be
    \bar \qr^{ZUJE} = \sum_{zuj}p_j \frac1{|\cU|}\ketbra{z,u,j}\otimes \sum_{xy} p_{xy} p_{z|ux} \bar\qr^E_{jxy}.
\ee
This is a sub-normalised state, with trace
\bea
    \tr\; \bar \qr^{ZUJE} &=& \tr_E \sum_j p_j \sum_{xy}p_{xy} P^j \qr^E_{jxy}  P^j
    \\ &=&
    \sum_j p_j \tr_E P^j \qr^E_j P^j
    \\ &=&
    \sum_j p_j \tr_E P^j \Big\{(1-\frac{3}{2} \gamma)\ketbra{0}+\frac{\gamma}{2} \sum_{k=1}^{3}\ketbra{k}\Big\}^{\otimes n} P^j
    \\ &=&
    \sum_j p_j \sum_{g\in\S_j} (\fr\qg2)^{w(g)}(1-\fr32\qg)^{n-w(g)}
    \\ &=&
    \sum_{(\qt_0,\qt_1,\qt_2,\qt_3)\in\cT} {n\choose \qt_0,\qt_1,\qt_2,\qt_3} \cdot  (1-\fr32\qg)^{\qt_0}
    (\fr\qg2)^{\qt_1+\qt_2+\qt_3}
%    \\ &=&
%    \sum_{(\qt_0,\qt_1,\qt_2)\in\cT} {n\choose \qt_0,\qt_1,\qt_2} (1-\fr32\qg)^{\qt_0} (\fr\qg2)^{\qt_1} \qg^{\qt_2}
\eea
Finally we use Lemma~\ref{lemma:distbound} to get 
$\|  \qr^{ZUJE}-\bar\qr^{ZUJE} \|_{\rm tr}\leq \sqrt{1\cdot(1-\tr \bar\qr^{ZUJE})}$.
\hfill$\square$

\begin{theorem}
\be
    \bar D \leq \fr12\sqrt{2^{\ell-n}} \sum_{(\qt_0,\qt_1,\qt_2,\qt_3)\in\cT} {n\choose \qt_0,\qt_1,\qt_2,\qt_3} 
    \left[\sqrt{(1-\qg)(1-\fr32\qg)}\right]^{\qt_0} 
    \left[\sqrt{\fr\qg2(1-\qg)}  \right]^{\qt_1} \left[\sqrt{\fr12 \qg^2} \right]^{\qt_2+\qt_3}.
\label{Dbartheorem}
\ee
\end{theorem}
\underline{\it Proof:}
We use
$P^j (\rho^E_{jx})^2  P^j - P^j \rho^E_{jx} P^j \rho^E_{jx} P^j $
$=P^j  \rho^E_{jx}   (\mathbb{I} - P^j) \rho^E_{jx}  P^j$
$=(P_\S \rho^E_{jx}  [\mathbb{I}-P^j])(P_\S \rho^E_{jx}  [\mathbb{I}-P^j])^\dagger$
$\geq 0$
to conclude that $(\bar\qr^E_{jx})^2 \leq P^j (\rho^E_{jx})^2  P^j$.
From Lemma~\ref{lemma:Adiag} we then get
$\sum_x p_x^2 (\bar\qr^E_{jx})^2$ 
$\leq \sum_{g\in\S_j}\ql_g(j)\ketbra{g}$.
Substitution into  (\ref{defDbar}) yields
\bea
    \bar D  &\leq& \fr12 \sqrt{2^{\ell}} \sum_j p_j \sum_{g\in\S_j}\sqrt{\ql_g(j)}
\eea
with the eigenvalues $\ql_g(j)$ as defined in (\ref{eigenvalueA}).
Since these eigenvalues depend only on the tallies, 
the sum over strings $g\in\S_j$ reduces to a sum over tallies in $\cT$
with multiplicity factor ${n\choose t_0,t_{\rm eq}, t_{+1}, t_{+2} }$.
Then, since the set $\cT$ has no dependence on $j$, the $\sum_j p_j$ reduces to~1.
\hfill$\square$

Note that (\ref{Dbartheorem}) can also be suggestively written as
\be
    \bar D \leq \fr12\sqrt{2^{\ell-n}} \sum_{(\qt_0,\qt_1,\qt_2,\qt_3)\in\cT} {n\choose \qt_0,\qt_1,\qt_2,\qt_3} 
    \sqrt{(1-\fr32\qg)^{\qt_0}(\fr\qg2)^{\qt_1+\qt_2+\qt_3} \cdot (1-\qg)^{\qt_0+\qt_1}\qg^{n-\qt_0-\qt_1}}
\label{Dbarsuggestive}
\ee
\be
    = \fr12\sqrt{2^{\ell-n}} \sum_{(\qt_0,\qt_1,\qt_2,\qt_3)\in\cT} {n\choose \qt_0,\qt_1,\qt_2,\qt_3} 
    (1-\fr32\qg)^{\qt_0}(\fr\qg2)^{\qt_1+\qt_2+\qt_3}
    \sqrt{\frac{(1-\qg)^{\qt_0+\qt_1}\qg^{n-\qt_0-\qt_1}}
    { (1-\fr32\qg)^{\qt_0}(\fr\qg2)^{\qt_1+\qt_2+\qt_3}  }}.
\label{Dbarsuggestive2}
\ee
The last line resembles an expectation of the square root expression, with a multinomial probability 
distribution. 

\begin{theorem}
\label{th:main}
Let $m=(m_0,m_1,m_2,m_3)\isdef(n[1-\fr32\qg] , n\fr\qg2, n\fr\qg2, n\fr\qg2)$.
Let $\cT$ be the set of tallies in an $\qa$-neighborhood of $m$, defined as
\be
    \cT = \{  (\qt_0,\qt_1,\qt_2,\qt_3)  \; | \; \qt_0+\qt_1+\qt_2+\qt_3=n \;\wedge\;
    \sum_{a=0}^3 |\qt_a-m_a| < \qa \sqrt n \}.
\ee
Then
\bea
    && \sum_{(\qt_0,\qt_1,\qt_2,\qt_3)\in\cT} {n\choose \qt_0,\qt_1,\qt_2,\qt_3} (1-\fr32\qg)^{\qt_0} (\fr\qg2)^{\qt_1+\qt_2+\qt_3}
    > 1- 16 e^{-\frac12 \qa^2}
\label{almostnorm}
	\\ &&
    \|  \qr^{ZUJE}-\bar\qr^{ZUJE} \|_{\rm tr}\leq 4 e^{-\frac14\qa^2}
\label{distexponential}
	\\ &&
    \bar D < \fr12\sqrt{2^{\ell-n}}\sqrt{2^{nh(1-\fr32\qg,\fr\qg2,\fr\qg2,\fr\qg2)-nh(\qg)}
    2^{\qa\sqrt n \fr12\log[\frac2\qg(1-\fr32\qg)]}
    }.
\label{thbarDwithalpha}
\eea
\end{theorem}
\underline{\it Proof:}
The summation in (\ref{epsilondist})
is a partial sum over a multinomial distribution which exactly matches the 
probability in Lemma~\ref{lemma:Bretagnolle}.
That proves (\ref{almostnorm}).
The upper bound (\ref{distexponential}) immediately follows.
Next, the summation in (\ref{Dbarsuggestive2}) can be interpreted 
(up to a factor $1-16e^{-\fr12\qa^2} < 1$) as an expectation of the square root
expression, for a multinomial distribution restricted to the set $\cT$.
We upper bound the expectation by the maximum attainable value on the set~$\cT$,
\be
    \bar D < \fr12\sqrt{2^{\ell-n}} \max_{(\qt_0,\qt_1,\qt_2,\qt_3)\in\cT} 
    \sqrt{\frac{(1-\qg)^{\qt_0+\qt_1}\qg^{n-\qt_0-\qt_1}}
    { (1-\fr32\qg)^{\qt_0}(\fr\qg2)^{\qt_1+\qt_2+\qt_3}  }}.
\label{maxwithalpha}
\ee
The fraction under the square root equals $[\frac{1-\qg}{1-\fr32\qg}]^{\qt_0}  [\fr2\qg(1-\qg)]^{\qt_1} 2^{\qt_2+\qt_3}$.
This is maximized by increasing $\qt_1$ as much as possible, at the cost of $\qt_0$, i.e.
$\qt_0=m_0-\fr12\qa\sqrt n$, $\qt_1=m_1+\fr12\qa\sqrt n$, $\qt_2=m_2$, $\qt_3=m_3$.
Substitution into (\ref{maxwithalpha}) yields (\ref{thbarDwithalpha}).
\hfill$\square$

%==================================================================
\section{Key rate}
\label{sec:keyrate}

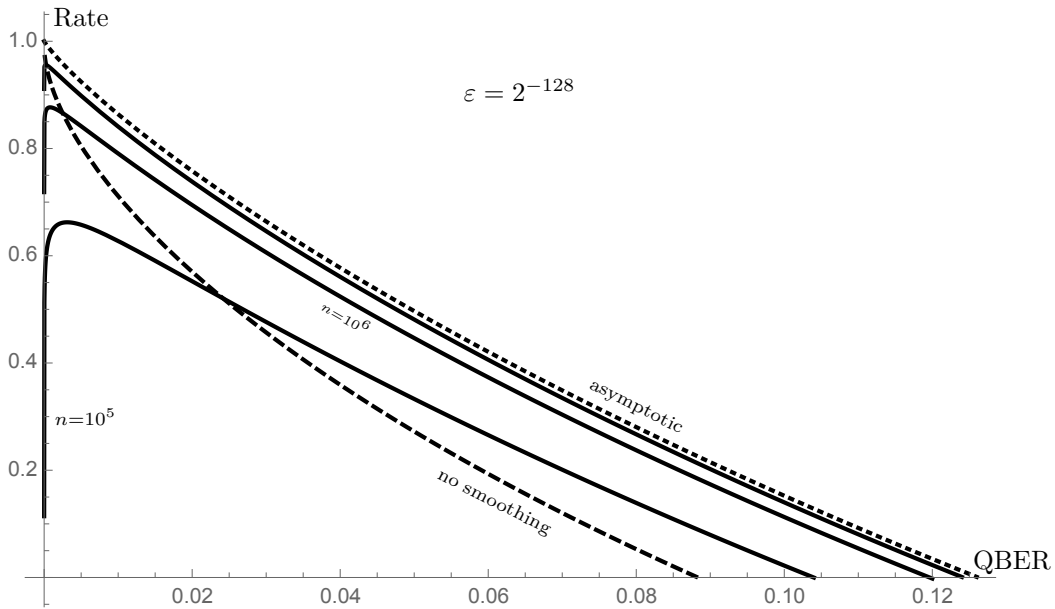
\begin{figure}[h]
\begin{center}
\setlength{\unitlength}{1mm}
\begin{picture}(140,80)(0,0)
\put(0,-3){\includegraphics[height=80mm]{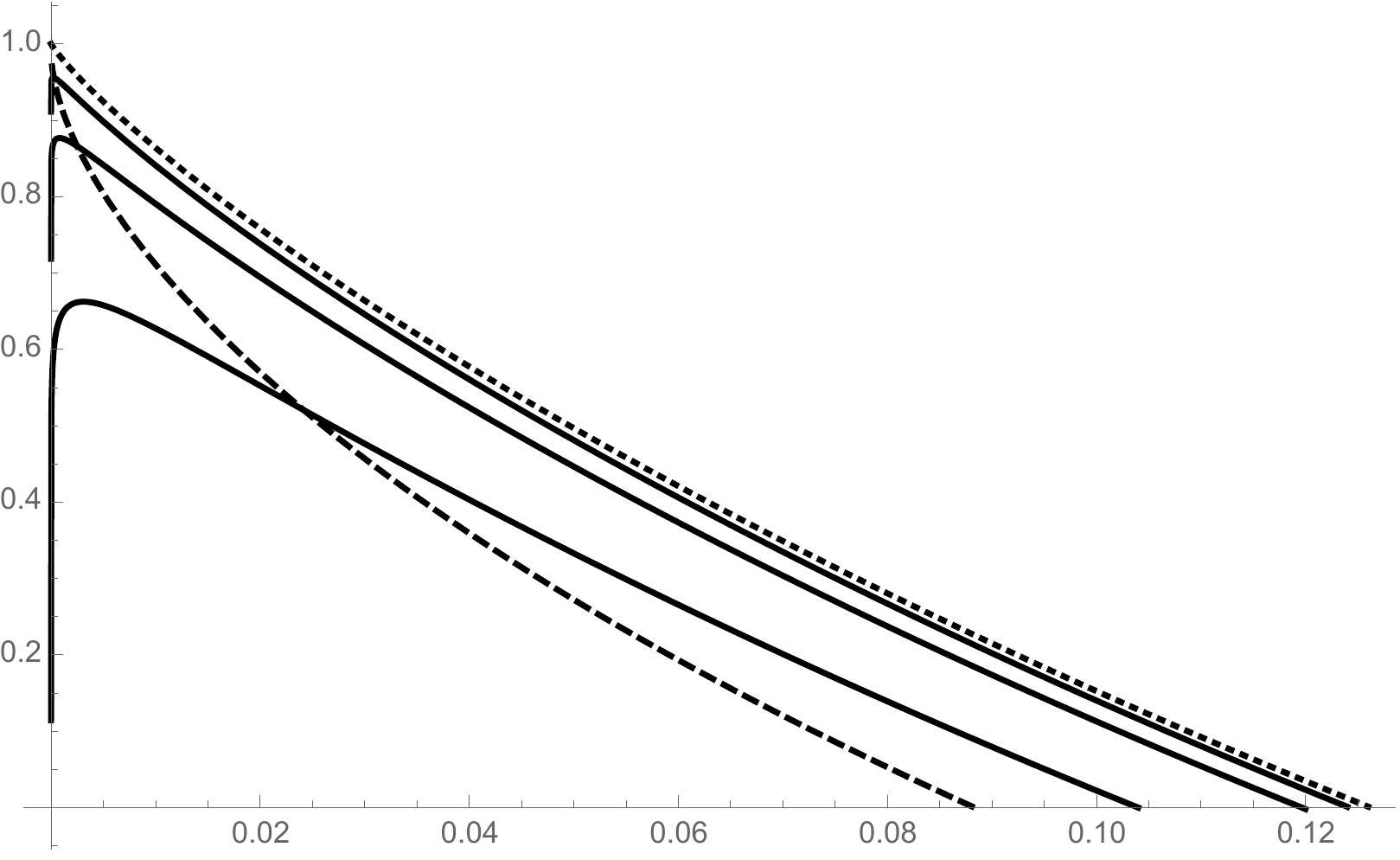}}
\put(6,75){Rate}
\put(127,3){QBER}
\put(60,65){$\qe=2^{-128}$}
\put(75,27){ \rotatebox{-27}   {${\scriptstyle {\rm asymptotic}}$}     }
\put(55,15){ \rotatebox{-27}   {${\scriptstyle {\rm no\; smoothing}}$}     }
\put(5,22){  ${\scriptstyle n=10^5}$ }
\put(40,38){\rotatebox{-30}{  ${\scriptscriptstyle n=10^6}$ }}
\end{picture}
\caption{\it 
Solid curves: The rate (\ref{therate}) as a function of the QBER ($\qg$) at $\qe=2^{-128}$,
plotted for $n=10^5$, $n=10^6$ and $n=10^7$. 
The dotted curve is the asymptotic rate (\ref{asymprate}).
The dashed curve is the rate obtained from the without-smoothing bound (\ref{Dnosmooth})
at $n=10^5$. 
}
\label{fig:rates}
\end{center}
\end{figure}

We discuss the key rate that follows from Theorem~\ref{th:main}.
Say that we want both $\bar D$ and the expression $\|  \qr^{ZUJE}-\bar\qr^{ZUJE} \|_{\rm tr}$
to be upper bounded by a constant~$\qe$.
Then according to (\ref{distexponential}) we need to set $\qa=2\sqrt{\ln (4/\qe)}$.
Substituting $\qa$ into (\ref{thbarDwithalpha}) we find that $\ell$ must be set to
\be
    \ell(\qe) =  n+2 - nh(1-\fr32\qg,\fr\qg2,\fr\qg2,\fr\qg2)+nh(\qg)
    - \sqrt{n \ln\fr4\qe} \cdot \log[\frac2\qg(1-\fr32\qg)]  - 2\log\frac1\qe.
\ee
The rate is obtained by subtracting from $\ell$ the size of the syndrome and the postselection penalty $30\log(n+1)$, 
and then normalising by a factor~$n$.
We assume the existence of an almost-perfect error correcting code,
such that the size of the syndrome is close to $nh(\qg)$.
\be
    {\rm Rate} \approx 1- h(1-\fr32\qg,\fr\qg2,\fr\qg2,\fr\qg2)
    -\frac1{\sqrt n} \sqrt{\ln\fr4\qe} \cdot \log[\frac2\qg(1-\fr32\qg)]
    - \frac{30\log n}{n} - \frac2n \log\frac1\qe.
\label{therate}
\ee
Note that 
(i) for $n\to\infty$ the asymptotic rate (\ref{asymprate}) is recovered;
(ii) leading-order finite size corrections of order $\sqrt{\fr1n \ln\fr1\qe}$ occur in the standard
proof technique too.
In Fig.\,\ref{fig:rates} we show how the obtained rate tends to the asymptotic result
as $n$ increases.

%==================================================================
\section{Discussion}
\label{sec:discussion}

The standard approach to smoothing departs from the Bell-diagonal structure of
$\sum_x p_x^2 (\qr^E_{jx})^2$.
We have shown that it is possible to get a good finite-size result
by retaining this structure.
It is interesting to note that our smoothing procedure 
is a simple restriction from full summation over $\G=\{0,1,2,3\}^n$
to the typical set $\S_j\subset\G$.
In contrast, the smoothing in \cite{RK2005} requires two different
operations, one to reduce a R\'{e}nyi-0 entropy
and one to increase a R\'{e}nyi-2 entropy.

We suggest a number of topics for future work.
(i) 
We did not try to get the sharpest possible bounds.
We expect that the constant in the $\cO(1/\sqrt n)$ finite-size contribution
can be reduced. In particular, the rate dip at small QBER may be avoided.
At low QBER one can just switch to the result without smoothing, but that is not very elegant.
(ii)
The proof method may be applied to other qubit-based schemes.

%\cite{KGR2005}
%\cite{SP2000}

%\cite{Mayers1996,BBBMR2000,Lo2001,GottesmanLo2003,Inamori2000,SP2000}

%%%%%%%%%%%%%%%%%%%%%%%%%%%%%%%%%%%%%%%%%%%%%%%%%%%%%%%%%%%%%%%%%%%%%%%%%%%%%%%%%%%%%%%

%-------------------------------------------------------------------------
%\section*{Acknowledgments}

\vskip3mm

{\large\bf Acknowledgments}\\
Part of this work was supported by NGF Quantum Delta NL KAT-2.

\appendix
\section{Appendix: Key Recycling}

%-----------------------------------------------------
\subsection{6-state Quantum Key Recycling}
\label{sec:prelimQKR}

In Quantum Key Recycling (QKR) \cite{bennett1983quantum,damgard2013quantum,fehr2017quantum,skoric2017quantum,leermakers2019security} 
the measurement bases are known beforehand, as part of a 
secret key shared by Alice and Bob.
In case of an {\em accept}, it is safe to re-use this secret.
Not having to discuss the measurement bases can eliminate one round of communication
between Alice and Bob.
Furthermore, there are no basis mismatches and hence no qubits have to be discarded.

A 6-state QKR scheme was studied in \cite{leermakers2019security}.
It encrypts an $\ell$-bit plaintext into a ciphertext that consists of $n$ qubits
and some classical data, including a one-time padded syndrome.
Part of the $\ell$-bit plaintext is reserved to carry the one-time pad for the next round.
The security analysis is very close to QKD.
A quantity $\bar D_{\rm qkr}$ similar to the trace distance $\bar D$ (\ref{defDbar}) needs to be made small.
It was shown that
$\bar D_{\rm qkr}\leq \fr12\sqrt{2^{\ell-n}} \tr\sqrt{\E_{jx}(\bar\qr^E_{jx})^2}$,
where $j$ is uniform.
Further analysis yields exactly the same R\'{e}nyi entropies as for QKD and the same asymptotic rate (\ref{asymprate}).
(The QKR rate is defined as the length of the actual message divided by the number of qubits).

It was noted in \cite{leermakers2019security} that the expression 
$\E_{jx} (\rho^E_{jx})^2$, i.e. without smoothing, is diagonal in the Bell basis.
This was exploited to obtain, without smoothing, a finite-size result for the QKR rate.
However, this rate is significantly lower than (\ref{asymprate}).

%---------------------------------------------------------------------

\subsection{Double smoothing}

The explicit-smoothing analysis for QKR is a bit more involved than for QKD.
The additional average over the basis choices $j\in\{1,2,3\}^n$
washes away the distinction between three of the tallies, and
allows for multiple values of the noise $r=\bar x\oplus y\in\bits^n$ to fit a string $g\in\G$,
whereas in QKD the $r$ is entirely fixed by~$g$.
Hence the eigenvalues of $\E_j \sum_x p_x^2 (\rho^E_{jx})^2$ involve an additional summation over $r$,
whose domain we need to restrict separately in order to get a good result for the rate.
This leads to a two-step smoothing procedure
that resembles the approach in \cite{RK2005}.
First we restrict summations over $r$ to a subset of Hamming weights $\cW\subset\{0,\ldots,n\}$.
We write the truncated version of $\qr^E_{jx}$ as $\qf^E_{jx}$,
\be
    \qf^E_{jx} 
    =\sum_{r:w(r)\in\cW} \mu_r  \bigotimes_{i=1}^n \qs^{j_i}_{x_i, \bar x_i\oplus r_i}
%    =\sum_{r:w(r)\in\cW} (1-\qg)^{n-w(r)}\qg^{w(r)} \bigotimes_{i=1}^n \qs^{j_i}_{x_i, \bar x_i\oplus r_i}
    \mbox{ with } \mu_r = (1-\qg)^{n-w(r)}\qg^{w(r)}.
\ee
Next we apply a projection $P_\S$ that restricts $\G$ to a subset $\S\subset \G$, but now not dependent on~$j$.
We get $\bar \qr^E_{jx} = P_\S \qf^E_{jx} P_\S$.
Next we bound $(\bar \qr^E_{jx})^2 \leq P_\S (\qf^E_{jx})^2 P_\S$, analogous to the QKD case,
to obtain 
$\E_j \sum_x p_x^2 (\bar\rho^E_{jx})^2\leq $
$P_\S \E_j\sum_x p_x^2 (\qf^E_{jx})^2 P_\S$.
We write $(\qf^E_{jx})^2 = \sum_{r:w(r)\in\cW} \mu_r^2  \bigotimes_{i=1}^n \qs^{j_i}_{x_i, \bar x_i\oplus r_i}$.
The averaged version of Lemma~\ref{sumsigmaisdiag} is
\be
    \E_{jx} \qs^j_{x,\bar x\oplus r} = \left\{  \begin{array}{ll} 
    r=0: &  \frac{1-\fr32\qg}{1-\qg}\ketbra{0} + \frac{\qg/6}{1-\qg}\sum_{j=1}^3\ketbra j  \\ 
    r=1: &  \fr13 \sum_{j=1}^3\ketbra j\end{array}\right.
\ee
This leads to a version of Lemma~\ref{lemma:Adiag} with different constants and different tallies,
\be
    P_\S \E_j \sum_x p_x^2 (\qf^E_{jx})^2 P_\S = \sum_{g\in\S} \qL_g \ketbra{g} 
\ee
with
\bea
    \qL_g &=& 2^{-n}\sum_{\stackrel{r: w(r)\in\cW}{g_i=0 \implies r_i=0}} \mu_r^2 
    (\frac{1-\fr32\qg}{1-\qg})^{t_0(g)}  (\frac13)^{w(r)}  (\frac{\qg/6}{1-\qg})^{n-t_0(g)-w(r)}
\label{rsumspecial}
    \\ &=&
    2^{-n} (1-\qg)^{t_0(g)}  (1-\fr32\qg)^{t_0(g)} (\frac\qg6)^{n-t_0(g)} \sum_{w\in\cW} {n-t_0(g)\choose w}
    (2\qg)^w (1-\qg)^{n-t_0(g)-w}.
\eea
The $r$-summation in (\ref{rsumspecial}) is restricted to those strings $r\in\bits^n$
that have $r_i=0$ in all locations~$i$ where $g_i=0$.
This leads to the combinatorial factor ${n-t_0\choose w}$.
Note that taking the full summation $\sum_{w=0}^{n-t_0}$ would reproduce the 
unsmoothed eigenvalues from \cite{leermakers2019security}.
Compared to the QKD proof, we need extra inequalities to bound the $w$-summation.
Let $w_{\rm min}$ be the lowest value in $\cW$.
We bound $\qL_g$ as
\be
    \qL_g < 2^{-n} (1-\fr32\qg)^{t_0}(\fr\qg2)^{n-t_0}(\fr13)^{n-t_0} |\cW|{n-t_0\choose w_{\rm min}}
    (2\qg)^{w_{\rm min}}(1-\qg)^{n-w_{\rm min}}
\ee
Using Stirling's approximation
$\sqrt{2\pi n}(\frac ne)^n e^{\frac1{12n+1}} < n! < \sqrt{2\pi n}(\frac ne)^n e^{\frac1{12n}}$
for the binomial we get
\bea
    \qL_g &<& 2^{-n} (1-\fr32\qg)^{t_0}(\fr\qg2)^{n-t_0}\qg^{w_{\rm min}}(1-\qg)^{n-w_{\rm min}}
    \nonumber\\ && \cdot
     \frac{|\cW|}{\sqrt{2\pi w_{\rm min}}} \cdot
    \frac{ 2^{w_{\rm min}} (\fr13)^{n-t_0} (1-\frac{t_0}n)^{n-t_0+\fr12} }
    { (\frac{w_{\rm min}}n)^{w_{\rm min}} (1-\frac{t_0}n-\frac{w_{\rm min}}n)^{n-t_0-w_{\rm min}+\fr12} }.
\label{stirlings}
\eea
Note that for $(t_0\approx n-n\fr32\qg, w_{\rm min}\approx n\qg) $
and $|\cW| \propto \sqrt{n\qg(1-\qg)}$ both fractions in (\ref{stirlings})
are almost constants.
Analogous to (\ref{Dbarsuggestive2}) we can obtain a bound
\bea
    \bar D_{\rm qkr} &<& \fr12\sqrt{2^{\ell-n}}\sum_{(\qt_0\cdots \qt_3)\in\cT}{n\choose \qt_0,\qt_1,\qt_2,\qt_3}
    (1-\fr32\qg)^{\qt_0}(\fr\qg2)^{\qt_1+\qt_2+\qt_3}
    \sqrt{   \frac{ \qg^{w_{\rm min}}(1-\qg)^{n-w_{\rm min}} }{ (1-\fr32\qg)^{\qt_0}(\fr\qg2)^{\qt_1+\qt_2+\qt_3} }        }
    \nonumber \\ && \cdot
    \sqrt{
     \frac{|\cW|}{\sqrt{2\pi w_{\rm min}}} \cdot
    \frac{ 2^{w_{\rm min}} (\fr13)^{n-\qt_0} (1-\frac{\qt_0}n)^{n-\qt_0+\fr12} }
    { (\frac{w_{\rm min}}n)^{w_{\rm min}} (1-\frac{\qt_0}n-\frac{w_{\rm min}}n)^{n-\qt_0-w_{\rm min}+\fr12} }
    }
\eea
which has the form of an incomplete multinomial expectation of the square root expression.
We can set $\cT$ as in Theorem~\ref{th:main} and similarly upper bound the mean by the maximum;
the maximum is again attained by setting $\qt_0=\qt_0^*\isdef n(1-\fr32\qg)-\fr12\qa\sqrt n$.
Thus the obtained bound is
\bea
    \bar D_{\rm qkr} &<& \fr12\sqrt{2^{\ell-n}}
    \sqrt{  2^{nh(1-\fr32\qg,\fr\qg2,\fr\qg2,\fr\qg2)}   \qg^{w_{\rm min}}(1-\qg)^{n-w_{\rm min}} 
      (\frac{\qg/2}{1-\fr32\qg})^{-\fr12 \qa\sqrt n} }
    \nonumber \\ && \cdot
    \sqrt{
     \frac{|\cW|}{\sqrt{2\pi w_{\rm min}}} \cdot
    \frac{ 2^{w_{\rm min}} (\fr13)^{n-\qt_0^*} (1-\frac{\qt_0^*}n)^{n-\qt_0^*+\fr12} }
    { (\frac{w_{\rm min}}n)^{w_{\rm min}} (1-\frac{\qt_0^*}n-\frac{w_{\rm min}}n)^{n-\qt_0^*-w_{\rm min}+\fr12} }
    }.
\eea
The asymptotic rate is the same as for QKD.

\bibliographystyle{plain} 
\bibliography{QKDsimple} 

\end{document}